\documentstyle[12pt]{article}

\newcommand{\be}{\begin{equation}}
\newcommand{\ee}{\end{equation}}
\makeatletter
\@addtoreset{equation}{section}
\makeatother

\newcommand{\complex}{{\kern .1em {\raise .47ex
\hbox {$\scriptscriptstyle |$}}
    \kern -.4em {\rm C}}}
\newcommand{\real}{{{\rm I} \kern -.19em {\rm R}}}
\newcommand{\rational}{{\kern .1em {\raise .47ex
\hbox{$\scripscriptstyle |$}}
    \kern -.35em {\rm Q}}}
\renewcommand{\natural}{{\vrule height 1.6ex width
.05em depth 0ex \kern -.35em {\rm N}}}
\newcommand{\dint}{\displaystyle{\int}}

\newcommand{\tr}{{\rm {Tr} \,}}

\newcommand{\fud}[2]  {{\displaystyle{\frac{\delta #1}{\delta #2}}}}
\newcommand{\dfrac}[2]{{\displaystyle{\frac{#1}{#2}}}}

\newcommand{\sla}{\raise.15ex\hbox{$/$}\kern -.57em}

\newcommand{\twiddle}{\lower.9ex\rlap{$\kern -.1em\scriptstyle\sim$}}

\newcommand{\equ}[1]{(\ref{#1})}

\newcommand{\eq}{\begin{equation}}
\newcommand{\eqn}[1]{\label{#1}\end{equation}}
\newcommand{\eea}{\end{eqnarray}}
\newcommand{\eqa}{\begin{eqnarray}}
\newcommand{\eqan}[1]{\label{#1}\end{eqnarray}}
\newcommand{\ba}{\begin{array}}
\newcommand{\ea}{\end{array}}
\newcommand{\eqac}{\begin{equation}\begin{array}{rcl}}
\newcommand{\eqacn}[1]{\end{array}\label{#1}\end{equation}}

\def\bea{\begin{eqnarray}}
\def\eea{\end{eqnarray}}

\def\bb#1{\hbox{\mybb#1}}
\def\complex{\bb{C}}

\def\real{\bb{R}}
\def\rational{\bb{Q}}
\def\R4{\bb{R}^4}
\font\mybb=msbm10 at 12pt

\def\tr{{\mbox{Tr}}}

\def\de{\partial}

\def\unita{{1 \kern-.30em 1}}
\begin{document}
\begin{titlepage}
\begin{flushright}
{ROM2F/97/08}\\
\end{flushright}
\vskip 1mm
\begin{center} 
{\large \bf Algebraic renormalization of the BF Yang-Mills Theory}\\   
\vspace{2.0cm}
{\bf F. Fucito}\footnote{{\sl I.N.F.N. Sezione di Roma II, 
Via Della Ricerca Scientifica, 00133 Roma, ITALY}},
{\bf M. Martellini}\footnote{Dipartimento di Fisica, 
Universit\`a di Milano, and I.N.F.N. Sezione di Milano, 
Via Celoria 16, 20133 Milano, ITALY}
\footnote{{\sl Landau Network  at ``Centro Volta'', Como,ITALY}},
{\bf S.P. Sorella}\footnote{{\sl UERJ, Universidade do 
Estado do Rio de Janeiro,Departamento de 
F\'\i sica Te\'orica,Instituto de F\'{\i}sica,Rua S\~ao 
Francisco Xavier, 524,20550-013, Marac
an\~{a}, Rio de Janeiro, Brazil}},\\
{\bf A. Tanzini}\hspace{1mm}$^{1\hspace{1mm}}$\footnote{{\sl 
Dipartimento di Fisica,
Universit\`a di Roma II ``Tor Vergata"}},
{\bf L.C.Q. Vilar}\footnote{{\sl C.B.P.F.,Centro Brasileiro 
de Pesquisas F\'\i sicas,Rua Xavier Sigaud 150,
22290-180 Urca,Rio de Janeiro, Brazil}}, 
{\bf M.Zeni}\hspace{1mm}$^{2\hspace{1mm}}$
\vskip 3.0cm
{\large \bf Abstract}\\
\end{center}
We discuss the quantum equivalence, to all orders of 
perturbation theory, between 
the Yang-Mills theory and its first order formulation through a second 
rank antisymmetric tensor field. Moreover, the introduction of an additional 
nonphysical vector field allows us to interpret the Yang-Mills theory as a 
kind of perturbation of the topological BF model. 
\vfill
\end{titlepage}
\addtolength{\baselineskip}{0.3\baselineskip}

\setcounter{section}{0}
\section{Introduction}
Among the various hypothesis proposed to explain the quark confinement, 
the description of the QCD vacuum as a dual magnetic 
superconductor is rather appealing 
\cite{mandel, thsc,poly}. 
Particularly noticeable is the formulation given by t'Hooft \cite{thooft1} who 
attempted to describe the vacuum of QCD by making use of an electric and a 
magnetic order parameter. The expectation values of these order
parameters can exhibit a perimeter or an area law, these different behaviours
labelling the various phases of QCD. However, in spite of the many efforts
devoted to put this description on firm mathematical basis, such an
achievement has not yet been satisfactorily accomplished. 
In order to improve such a situation, some of the present authors have studied
an alternative first order formulation of the non-Abelian Yang-Mills (YM) gauge
theory \cite{fmz,pcr}. This  formulation (called from now on BF-YM) makes use 
of an anti-symmetric tensor $B_{\mu\nu}$ and of an additional vector 
field $\eta_{\mu}$, yielding the following classical action 
\eq
I_{BF-YM}=-\int \tr [B\wedge F + g^2(B-D\eta)\wedge *(B-D\eta)],
\eqn{class-action}
where $F=dA+A\wedge A$ is the YM field strength and $(D\eta=d\eta + 
[A,\eta])$ the covariant derivative. All the fields are Lie algebra
valued, 
the generators $T^a$ of the corresponding gauge group G being choosen to be
antihermitians and normalized in the fundamental representation
as $\tr(T^aT^b)=-\delta^{ab}/2$. 
The Hodge dual of a $p$-form in $D$ dimension is defined as
$*=\varepsilon^{\mu_1\ldots\mu_D}/(D-p)!$, and the exterior product between a
$p$ form $\omega$ and a $q$ form $\xi$ is given by 
\eq
\omega\wedge\xi={1\over p!q!}\varepsilon^{\mu_1\ldots\mu_{p+q}}
\omega_{\mu_1\ldots\mu_p}\xi_{\mu_{p+1}\ldots\mu_{p+q}}d^{p+q}x.
\eqn{wedge}
It is very easy to see that if one eliminates the antisymmetric field $B$ from
the action \equ{class-action} using the equations of motion
\eq\ba{l}
 \dfrac{1}{2}D_\nu *B^{\mu\nu} =
g^2[\eta_\sigma,B^{\mu\sigma}-D^{[\mu}\eta^{\sigma]}],                                                                   \\[3mm]
\dfrac{1}{2} *F^{\mu\nu} = -g^2(B^{\mu\nu}-D^{[\mu}\eta^{\nu]}),   \\[3mm]
D_\rho B^{\rho\nu} = (D^2g^{\mu\nu}-D^\mu D^\nu)\eta_\mu ,  
\ea\eqn{eq-mot}
one recovers the usual form of the YM action. 
Moreover, as discussed in \cite{fmz}, the tensor field $B$ allows to 
obtain an explicit expression for the magnetic order parameter whose 
commutation relation with the electric order parameter gives 
the correct result and whose
expectation value turns out to obey the desired perimeter law \cite{fmz}. 
Furthermore, once acting on physical states, this magnetic operator gives a
singular gauge transformation, as it should be. 
Having clarified the role of the tensor field $B$, let us now spend a few words
about the 
additional vector field $\eta_{\mu}$  present in the expression
\equ{class-action} \cite{catt}. In order to motivate its introduction, let us first observe
that in the limit of zero coupling constant the action \equ{class-action}
reduces to the topological $BF$ action \cite{bbrt} which, in addition to the
gauge invariance, is known to possess a further local 
tensorial invariance whose origin is deeply related to the topological
character of the 
$BF$ system. Of course, by adding to the pure $BF$ only the term  
$B\wedge *B$ one always recovers the usual YM action, but the topological tensor
symmetry is lost. However, as one can easily understand, the
introduction of the vector field $\eta_{\mu}$ provides a simple way to
compensate the breaking induced by the term $B\wedge *B$, restoring thus the
topological tensor invariance. In other words the action \equ{class-action},
although classically equivalent to the YM theory, preserves all the simmetries
of the topological $BF$ model, giving us the interesting possibility of looking
at the pure YM as a perturbation of a topological model \cite{noi}. Let also
remark that, as we shall see in the next section, the transformation 
law of the vector field $\eta_{\mu}$ is simply given by a shift,
meaning that all the components of $\eta_{\mu}$ are nonphysical. It is worth to
notice that such a similar vector field has been recently used \cite{niemi} in
order to implement an alternative Higgs mechanism in which the YM gauge fields
acquire a mass through the breaking of a topological symmetry. This is another
rather attractive aspect related to the action \equ{class-action}. 
Let us focus, for the time being, on the aim of this letter, {\it i.e.} on the
study of the quantization and the renormalizability of \equ{class-action}  as
well as of the 
proof of its quantum equivalence with the ordinary YM theory, this 
being a necessary first consistency check supporting the usefulness of the action
\equ{class-action}. In particular, we shall be able to give a complete
algebraic proof of the quantum equivalence of \equ{class-action} with YM based
on BRST cohomological arguments. We emphasize here that such
an algebraic proof extends to all orders of perturbation theory and does not
rely on the existence of a regularization preserving the symmetries, 
being particularly adapted to the present case 
due to the presence of the
Levi-Civita tensor. Finally, let us mention that the study of the quantum
equivalence has been recently discussed in three dimensions by \cite{bf3}, 
and in four dimensions, using different techniques, by \cite{noi}. 
The work is organized as follows. In Sect.2 we analyse the symmetry
content of the action \equ{class-action} and we establish the classical
Slavnov-Taylor identity. Sect.3 is devoted to the study of the quantum aspects.

\section{Quantization of BF-YM theory}
The action \equ{class-action} is easily seen to be left invariant by the
following 
transformations 
\eq\ba{l}
\delta A_\mu = \delta_G A_\mu +\delta_T A_\mu+\delta^\prime A_\mu,  \\
\delta B_{\mu\nu} = \delta_G B_{\mu\nu}+\delta_T B_{\mu\nu} +\delta^\prime
B_{\mu\nu}, \\
\delta \eta_\mu = \delta_G \eta+\delta_T \eta_\mu+
  \delta^\prime \eta_\mu, 
\ea\eqn{class-symm}
where  $\delta_G$ and $\delta_T$ denote respectively the generators of the
gauge and of the tensorial topological invariance defined by 
\eq\ba{l} 
\delta_G A_{\mu} =  D_\mu\theta, \\
\delta_G B_{\mu\nu}=  [B_{\mu\nu},\theta], \\
\delta_G\eta_\mu=  [\eta_\mu,\theta],
\ea\eqn{gauge-inv}
and 
\eq\ba{l}
\delta_T A_{\mu}  = 0, \\
\delta_T B_{\mu\nu} =  D_{[\mu}\epsilon_{\nu]}, \\
\delta_T\eta_\mu  =  \epsilon_\mu. 
\ea\eqn{top-inv}
In particular, from the eqs.\equ{top-inv} one can see that the 
topological tensor transformation of the vector field $\eta_\mu$ is given by 
a shift, meaning that all the components of $\eta_\mu$ are nonphysical.  
The third generator $\delta^\prime$ appearing in eqs.\equ{class-symm} is
associated to a further local invariance whose transformations are given by 
\eq\ba{l}
\delta^\prime A_\mu=0, \\
\delta^\prime B_{\mu\nu}=[F_{\mu\nu},\sigma],\\
\delta^\prime \eta_\mu=D_\mu\sigma.
\ea\eqn{furth-inv}
We remark that the symmetry \equ{class-symm} is reducible since $\delta_T$ and
$\delta^\prime$ are not independent, as it can be seen by choosing
$\epsilon_\mu=D_\mu\sigma$. 
In order to gauge fix the local invariance \equ{class-symm} of the action 
\equ{class-action} we adopt the linear gauge conditions
\eq
\de^\mu A_\mu=0, \qquad \de^\mu B_{\mu\nu}=0, \qquad \de^\mu \eta_\mu=0. 
\eqn{gauge-cond}
Following the BRST formalism \cite{brs}, the gauge fixing action in a Landau
type gauge is then given by 
\eq
I_{gf}=\int d^4x\: s\tr\left[ \bar c \de^\mu A_\mu +\bar\psi^\nu\de^\mu
B_{\mu\nu}+\bar\rho\de^\mu \eta_\mu 
  +(\de^\mu\bar\psi_\mu)u+\bar\phi\de^\mu \psi_\mu\right], 
\eqn{gauge-fix-term}
where $(c, \bar c, h_A)$, $(\psi, \bar\psi, h_B)$, $(\rho, \bar\rho, h_\eta)$
are respectively the ghost, the antighost and the lagrangian multiplier for  
$\delta_G$, $\delta_T$, $\delta^\prime$; $(\phi, \bar\phi, h_\psi)$
the ghost, the antighost and the lagrangian multiplier for the zero modes of
the topological symmetry $\delta_T$, and $(u, h_{\bar\psi})$ a pair of fields
which takes into account a further degeneracy associated with $\bar\psi$.
The dimensions and the ghost numbers of all the fields are summarized in Table
1.
\begin{center}
\begin{tabular}{|c||c|c|c|c|c|c|c|c|c|}
\hline Fields & A & B & $\eta$ & c & $\bar c$ & $\psi$ & $\bar\psi$ & $h_A$ &
$h_B$ \\ \hline\hline dimension &  1 & 2 & 1 & 0 & 2 & 1 & 1 & 2 & 1 \\ 
\hline ghost \# & 0 & 0 & 0 & 1 & -1 & 1 & -1 & 0 & 0 \\ 
\hline
\end{tabular}\\
\vspace{.5cm}
\begin{tabular}{|c||c|c|c|c|c|c|c|c|}
\hline Fields & $\phi$ & $\bar\phi$ & $h_\psi$ & $\rho$ & $\bar\rho$ & $h_\eta$
& u & $h_{\bar\psi}$ \\ 
\hline\hline dimension &  0 & 2 & 2 & 0 & 2 & 2 & 2 & 2  \\ 
\hline ghost \# & 2 & -2 & -1 & 1 & -1 & 0 & 0 & 1  \\ 
\hline
\end{tabular}\\
\end{center}
\vspace{3mm}
\centerline{ Table 1}
The BRST transformations of the fields are
\eq\ba{l}
sA_\mu = -D_\mu c,  \\
sB_{\mu\nu} = -[B_{\mu\nu},c]+D_{[\mu}\psi_{\nu]}+ [F_{\mu\nu},\rho],\\
s\eta_\mu = -[\eta_\mu,c]+\psi_\mu+D_\mu\rho, \\
sc = {1\over 2}[c,c],\qquad s\bar c=h_A, \\
s\psi_\mu = [\psi_\mu,c]+D_\mu\phi,\qquad s\bar\psi_\mu=h_B, \\
sh_A = 0,\qquad sh_B=0,\qquad sh_\psi=0,\qquad sh_\eta=0, \\
s\phi =-[\phi,c], \qquad s\bar\phi=h_\psi,\\
s\rho = [\rho,c]-\phi, \qquad s\bar\rho=h_\eta, \\
su = h_{\bar\psi},\qquad sh_{\bar\psi}=0,
\ea\eqn{BRST-transf}
where the parenthesis $[\cdot,\cdot]$ denotes the graded commutator. It is easy
to see 
that $[s,s]=0$ on all the fields, {\it i.e.} the operator $s$ is nilpotent
off-shell.
In order to write down a Slavnov-Taylor identity corresponding to the
transformations \equ{BRST-transf} we introduce a set of external sources
($A^\ast_\mu,B^\ast_{\mu\nu},\eta^\ast_\mu,\psi^\ast_\mu,\rho^\ast,c^\ast,
\phi^\ast$)
coupled to the non-linear variations of eqs.\equ{BRST-transf}
\eq\ba{l}
 I_{ext}=\tr \dint d^4x\:\bigg(-A^{\ast}_\mu D^\mu
c+{1\over2}B^{\ast}_{\mu\nu}
        (D^{[\mu}\psi^{\nu]}-[B^{\mu\nu},c]+[F^{\mu\nu},\rho]) \\
\qquad \qquad \qquad \qquad
+\eta_{\mu}^{\ast}(\psi^\mu-[\eta^\mu,c]+D^\mu\rho)+ \psi_{\mu}^{\ast}
 (D^\mu \phi+[\psi^\mu ,c]) \\
\qquad \qquad \qquad \qquad +\rho^\ast(-\phi+[\rho,c])+{1\over2}c^\ast[c,c]
        -\phi^\ast[\phi,c]\bigg).
\ea\eqn{ext-action}
Therefore, the complete action 
\eq
\Sigma= I_{BF-YM}+I_{gf}+I_{ext},
\eqn{comp-actiomn}
satisfies the Slavnov-Taylor identity
\eq
{\cal S}(\Sigma)=0,
\eqn{slav-Tayl-id}
where
\eq\ba{l}
{\cal S}(\Sigma)=\tr \dint d^4x\:\bigg( 
     {\fud{\Sigma}{A^\mu}}{\fud{\Sigma}{A^\ast_\mu}}
   + {1\over 2}{\fud{\Sigma}{B^{\mu\nu}}}{\fud{\Sigma}{B^\ast_{\mu\nu}}}  
   + {\fud{\Sigma}{\eta^\mu}}{\fud{\Sigma}{\eta^\ast_\mu}}
   + {\fud{\Sigma}{\psi^\mu}}{\fud{\Sigma}{\psi^\ast_\mu}}
   + {\fud{\Sigma}{\rho}}{\fud{\Sigma}{\rho^\ast}}  \\[3mm]
     \qquad \qquad \qquad
   + {\fud{\Sigma}{ c}}{\fud{\Sigma}{ c^\ast}}
   + {\fud{\Sigma}{\phi}}{\fud{\Sigma}{\phi^\ast}}
   + h_A{\fud{\Sigma}{\bar c}}    + h^\mu_B{\fud{\Sigma}{\bar\psi^\mu}}
   + h_{\bar\psi}{\fud{\Sigma}{ u}}+ h_\psi{\fud{\Sigma}{\bar\phi}}
   + h_\eta{\fud{\Sigma}{\bar\rho}}\bigg).
\ea\eqn{slav-tayl-op}
In addition to the Slavnov-Taylor identity \equ{slav-Tayl-id}, the complete
action $\Sigma$ turns out to be characterized by the following additional
constraints:  
\begin{itemize}
\item the Landau gauge fixing conditions
\eq\ba{l}
{\fud{\Sigma}{h_A}}=\de^\mu A_\mu, \qquad \qquad 
{\fud{\Sigma}{h_B^\nu}}=\de^\mu B_{\mu\nu}-\de_\nu u, \\[3mm]
{\fud{\Sigma}{h_\eta}}=\de^\mu\eta_\mu, \qquad \qquad 
{\fud{\Sigma}{h_\psi}}=\de^\mu\psi_\mu, \\[3mm]
{\fud{\Sigma}{h_{\bar\psi}}}=\de^\mu\bar\psi_\mu, \qquad \qquad 
{\fud{\Sigma}{ u}}=\de^\mu h_{B\mu}; 
\ea\eqn{gauge-fix-cond}
\item the antighost equations following from the Slavnov-Taylor identity
\equ{slav-Tayl-id} and the gauge conditions \equ{gauge-fix-cond}
\eq\ba{l}
{\fud{\Sigma}{\bar c}}+ \de^\mu{\fud{\Sigma}{A^\ast_\mu}}=0, 
   \qquad \qquad 
{\fud{\Sigma}{\bar\psi_\nu}}+ \de^\mu{\fud{\Sigma}{B^\ast_{\mu\nu}}}=\de_\nu
h_{\bar\psi},     \\[3mm]
{\fud{\Sigma}{\bar\rho}}+ \de^\mu{\fud{\Sigma}{\eta^\ast_\mu}}=0, 
   \qquad \qquad 
{\fud{\Sigma}{\bar\phi}}- \de^\mu{\fud{\Sigma}{\psi^\ast_\mu}}=0;
\ea\eqn{antigh-eqs}
\item the ghost equation, usually valid in the Landau gauge \cite{ps} 
\eq
\dint d^4x\:\bigg(
 {\fud{\Sigma}{c}} +\big[{\bar c,{\fud{\Sigma}{h_A}}}\big]
 + \big[{\bar\psi^\nu,{\fud{\Sigma}{h_B^\nu}}}\big]
 + \big[{u,{\fud{\Sigma}{h_{\bar\psi}}}}\big] 
 + \big[{\bar\rho,{\fud{\Sigma}{h_\eta}}}\big]
 + \big[{\bar\phi,{\fud{\Sigma}{h_\psi}}}\big]\bigg)=\Delta_{cl}^c, 
\eqn{ghost-eq}
where $\Delta_{cl}^c$ is a linear classical breaking given by 
\eq\ba{l}
\Delta_{cl}^c=\dint d^4x\:\tr\bigg(-[A^\ast_\mu,A^\mu] + 
   {1\over2}[B^\ast_{\mu\nu},B^{\mu\nu}]+
 [\eta_\mu^\ast,\eta^\mu]-[\psi^\ast_\mu,\psi^\mu] \\
\qquad \qquad \qquad \qquad
-[\rho^\ast,\rho]+[\phi^\ast,\phi]-[c^\ast,c]\bigg);
\ea\eqn{gh-class-break}
\item the Ward identity for the rigid gauge invariance stemming from the
Slavnov-Taylor identity \equ{slav-Tayl-id} and the ghost equation
\equ{ghost-eq} 
\eq
{\cal H}_{rig}\Sigma=\sum_{all~fields~\Phi}\int
d^4x\:\left[{\Phi,{{\delta\Sigma}\over{\delta\Phi}}}\right]=0;
\eqn{rig-inv}
\item the linearly broken Ward identity corresponding to the ghost $\phi$,
typically of a topological $BF$ system \cite{lps}
\eq
\dint d^4x\:\bigg({{{\delta\Sigma}\over{\delta\phi}} 
 - \big[{\bar\phi ,{{\delta\Sigma}\over{\delta
h_A}}}\big]}\bigg)=\Delta_{cl}^\phi,
\eqn{phi-eq}
where $\Delta_{cl}^\phi$ is given by
\eq
\Delta_{cl}^\phi = \dint
d^4x\:\tr\bigg({[\psi^\ast_\mu,A^\mu]-\rho^\ast+[\phi^\ast,c]}\bigg);
\eqn{phi-class-break}
\item the Ward identity following from the $\phi$ ghost equation \equ{phi-eq}
and the Slavnov-Taylor identity \equ{slav-Tayl-id}
\eq
\dint d^4x\:\bigg({{\delta\Sigma}\over{\delta\rho}}
 + \big[{A^\mu,{{\delta\Sigma}\over{\delta\psi_\mu}}}\big]
 + \big[{c, {{\delta\Sigma}\over{\delta\phi}}}\big]
 - \big[{\psi^\ast_\mu,{{\delta\Sigma}\over{\delta A_\mu^\ast}}}\big]
 + \big[{\phi^\ast,{{\delta\Sigma}\over{\delta c^\ast}}}\big]
 + \big[{\bar\phi,{{\delta\Sigma}\over{\delta\bar c}}}\big]
 - \big[{h_\psi,{{\delta\Sigma}\over{\delta h_A}}}\big]\bigg)=0.
\eqn{rho-ward-id}
\end{itemize}
Notice, finally, that the breaking terms in the left hand side of the equations
\equ{ghost-eq} and \equ{phi-eq}, being linear in the quantum fields, are
classical 
breakings, {\it i.e.} they are present only at the classical level and will not
get renormalized by the radiative corrections \cite{ps}.


\section{Renormalization and algebraic equivalence with YM theory}

We face now the problem of the quantum extension of the Slavnov-Taylor identity
\equ{slav-Tayl-id} and of the equivalence with the YM theory. By following
standard arguments \cite{ps}, all the constraints
\equ{gauge-fix-cond}--\equ{rho-ward-id} derived in the previous  section 
can be shown to be renormalizable. They can therefore be assumed to hold 
for the quantum
vertex functional
\eq
\Gamma=\Sigma + O(h).
\eqn{quantum-action}
In particular, the gauge conditions \equ{gauge-fix-cond} imply that the higher
order contributions to the vertex functional $\Gamma$ are independent from
the lagrangian multipliers and that, due to the  equations
\equ{antigh-eqs}, the antighosts $({\bar c}, {\bar \psi}, {\bar \rho}, {\bar
\phi})$ enter only through the combinations
\bea
&&\hat A^\ast_\mu=A^\ast_\mu+\de_\mu\bar c,\nonumber\\
&&\hat B^\ast_{\mu\nu}=B^\ast_{\mu\nu}+\de_{[\mu}\bar\psi_{\nu]},\nonumber\\
&&\hat\eta^\ast_\mu=\eta^\ast_\mu+\de_\mu\bar\rho,\nonumber\\
&&\hat\psi^\ast_\mu=\psi^\ast_\mu-\de_\mu\bar\phi.
\label{shifted-antif}
\eea
Introducing then the reduced action $\hat\Sigma$ \cite{ps}
\bea
&&\hat\Sigma=I_{BF-YM}+\int d^4x\:\tr\bigg(-\hat A^{\ast}_\mu D^\mu c+
{1\over 2}\hat B^{\ast}_{\mu\nu}(D^{[\mu}\psi^{\nu]}-
[B^{\mu\nu},c]+[F^{\mu\nu},\rho])\nonumber\\
&& \qquad \qquad \qquad \qquad
+\hat \eta_{\mu}^{\ast}(\psi^\mu-[\eta^\mu,c]+D^\mu\rho)+
\hat \psi_{\mu}^{\ast}(D^\mu \phi+[\psi^\mu ,c])\nonumber\\
&& \qquad \qquad \qquad \qquad +\rho^\ast (-\phi+[\rho,c])+{1\over
2}c^\ast[c,c]-
\phi^\ast[\phi,c]\bigg),
\label{red-action}
\eea
the Slavnov-Taylor identity \equ{slav-Tayl-id} takes the following simpler form
\be
{\cal B}_{\hat\Sigma}\hat\Sigma=0.
\label{due.4}
\ee
where the operator ${\cal B}_{\hat\Sigma}$ is defined as
\bea
&&{\cal B}_{\hat\Sigma}=\int d^4x\:\tr\bigg({{\delta\hat\Sigma}\over{\delta
A^\mu}}
{\delta\over{\delta\hat A^\ast_\mu}}+
{{\delta\hat\Sigma}\over{\delta\hat A^\ast_\mu}}{\delta\over A^\mu}+
{{\delta\hat\Sigma}\over{\delta\eta^\mu}}{\delta\over{\delta\hat\eta^\ast_\mu}}+
{{\delta\hat\Sigma}\over{\delta\hat\eta^\ast_\mu}}
{\delta\over{\delta\eta^\mu}}+{1\over 2}{{\delta\hat\Sigma}\over{\delta
B_{\mu\nu}}}
{\delta\over{\delta\hat B^\ast_{\mu\nu}}} \nonumber\\
&& \qquad \qquad \qquad +{1\over 2}
{{\delta\hat\Sigma}\over{\delta\hat B^\ast_{\mu\nu}}}
{\delta\over{\delta B_{\mu\nu}}}+
{{\delta\hat\Sigma}\over{\delta\psi_\mu}}
{\delta\over{\delta\hat\psi^\ast_\mu}}+
{{\delta\hat\Sigma}\over{\delta\hat\psi^\ast_\mu}}
{\delta\over{\delta\psi_\mu}}+
{{\delta\hat\Sigma}\over{\delta\rho}}
{\delta\over{\delta\rho^\ast}}+
{{\delta\hat\Sigma}\over{\delta\rho^\ast}}
{\delta\over{\delta\rho}} \nonumber\\
&& \qquad \qquad \qquad +{{\delta\hat\Sigma}\over{\delta\phi}}
{\delta\over{\delta\phi^\ast}}+
{{\delta\hat\Sigma}\over{\delta\phi^\ast}}
{\delta\over{\delta\phi}}+
{{\delta\hat\Sigma}\over{\delta c}}
{\delta\over{\delta c^\ast}}+
{{\delta\hat\Sigma}\over{\delta c^\ast}}
{\delta\over{\delta c}}\bigg),
\label{lin-slav-tayl-op}
\eea
with
\eq
 {\cal B}_{\hat\Sigma}{\cal B}_{\hat\Sigma} = 0.
\eqn{nilp}
Its action on the fields and on the sources is given by
\bea
{\cal B}_{\hat\Sigma}A_\mu&=&-D_\mu c,\nonumber\\
{\cal B}_{\hat\Sigma}B_{\mu\nu}&=&-[B_{\mu\nu},c]+D_{[\mu}\psi_{\nu]}+
[F_{\mu\nu},\rho],\nonumber\\
{\cal B}_{\hat\Sigma}\eta_\mu&=&-[\eta_\mu,c]+\psi_\mu+D_\mu\rho,\nonumber\\
{\cal B}_{\hat\Sigma}c&=&{1\over 2}[c,c],\nonumber\\
{\cal B}_{\hat\Sigma}\psi_\mu&=&[\psi_\mu,c]+D_\mu\phi,\nonumber\\
{\cal B}_{\hat\Sigma}\phi&=&-[\phi,c],\nonumber\\
{\cal B}_{\hat\Sigma}\rho&=&[\rho,c]-\phi,\nonumber\\
{\cal B}_{\hat\Sigma}\hat A^\ast_\mu&=&{1\over 2}\varepsilon_{\mu\nu\rho\sigma}
D^\nu B^{\rho\sigma}+2g^2[\eta^\sigma, B_{\mu\sigma}-D_{[\mu}\eta_{\sigma]}]+
[c,\hat A_\mu^\ast]-[\rho,\hat\eta^\ast_\mu]+\nonumber\\
&&-[\psi^\nu,\hat B^\ast_{\mu\nu}]+D^\nu[\hat B^\ast_{\nu\mu},\rho]+
[\phi,\hat\psi^\ast_\mu],\nonumber\\
{\cal B}_{\hat\Sigma}\hat B^\ast_{\mu\nu}&=&{1\over 2}
\varepsilon_{\mu\nu\rho\sigma}F^{\rho\sigma}+
2g^2(B_{\mu\nu}-D_{[\mu}\eta_{\nu]})+[c,\hat B^\ast_{\mu\nu}],\nonumber\\
{\cal B}_{\hat\Sigma}\hat\eta^\ast_\mu&=&-2g^2D^\nu B_{\mu\nu}
-2g^2(D^2g_{\mu\nu}-D_\nu D_\mu)\eta^\nu+[c,\hat\eta^\ast_\mu],\nonumber\\
{\cal B}_{\hat\Sigma}\hat\psi^\ast_\mu&=&[c,\hat\psi^\ast_\mu]
-D^\nu\hat B^\ast_{\mu\nu}-\hat\eta^\ast_\mu,\nonumber\\
{\cal B}_{\hat\Sigma}\rho^\ast&=&{1\over 2}[F^{\mu\nu},\hat B^\ast_{\mu\nu}]
+[c,\rho^\ast]+D^\mu\hat\eta^\ast_\mu,\nonumber\\
{\cal B}_{\hat\Sigma}c^\ast&=&-D^\mu\hat
A^\ast_\mu+[\hat\eta^\ast_\mu,\eta^\mu]
+{1\over 2}[\hat B^\ast_{\mu\nu},B^{\mu\nu}]
-[\hat\psi^\ast_\mu,\psi^\mu]+\nonumber\\
&&+[\rho,\rho^\ast]+[\phi^\ast,\phi]+[c,c^\ast],\nonumber\\
{\cal
B}_{\hat\Sigma}\phi^\ast&=&-\rho^\ast+[c,\phi^\ast]-D^\mu\hat\psi^\ast_\mu.
\label{lin-op-action}
\eea
As it is well known, both the anomalies and the invariant counterterms
can be characterized as nontrivial cohomology classes of the operator
${\cal B}_{\hat\Sigma}$, {\it i.e.} they are solution of the consistency
condition
\be
{\cal B}_{\hat\Sigma}\Delta=0,
\label{due.7}
\ee
$\Delta$ being a local integrated polynomial of canonical dimension $4$
and ghost number $0$ for the counterterms and $1$ for the anomalies.
\noindent
In order to compute the cohomology
of
${\cal B}_{\hat\Sigma}$ we begin by analysing the cohomology of the operator
${\cal B}_{\hat\Sigma}^{(0)}$, corresponding to the linearized approximation of
eqs.(\ref{lin-op-action}), {\it i.e.}
\bea
{\cal B}_{\hat\Sigma}^{(0)} A_\mu&=&-\de_\mu c,\nonumber\\
{\cal B}_{\hat\Sigma}^{(0)} B_{\mu\nu}&=& \de_{[\mu}\psi_{\nu]},\nonumber\\
{\cal B}_{\hat\Sigma}^{(0)} \eta_\mu&=&\psi_\mu+\de_\mu\rho,\nonumber\\
{\cal B}_{\hat\Sigma}^{(0)} \psi_\mu&=&\de_\mu\phi,\nonumber\\
{\cal B}_{\hat\Sigma}^{(0)} \phi&=&0,\nonumber\\
{\cal B}_{\hat\Sigma}^{(0)} c&=&0,\nonumber\\
{\cal B}_{\hat\Sigma}^{(0)} \rho&=&-\phi,\nonumber\\
{\cal B}_{\hat\Sigma}^{(0)}\hat A^\ast_\mu&=&{1\over 2}
 \varepsilon_{\mu\nu\rho\sigma}\de^\nu B^{\rho\sigma},\nonumber\\
{\cal B}_{\hat\Sigma}^{(0)}\hat B^\ast_{\mu\nu}&=& {1\over 2}
 \varepsilon_{\mu\nu\rho\sigma}\de^{[\rho}A^{\sigma]}
+2g^2(B_{\mu\nu}-\de_{[\mu}\eta_{\nu]}),\nonumber\\
{\cal B}_{\hat\Sigma}^{(0)}\hat\eta^\ast_\mu&=&-2g^2(\de^\nu B_{\mu\nu}
+\de^2\eta_\mu-\de^\nu\de_\mu\eta_\nu),\nonumber\\
{\cal B}_{\hat\Sigma}^{(0)}\hat\psi^\ast_\mu&=&-\de^\nu\hat B^\ast_{\mu\nu}
-\hat\eta^\ast_\mu,\nonumber\\
{\cal
B}_{\hat\Sigma}^{(0)}\phi^\ast&=&-\rho^\ast-\de^\mu\hat\psi^\ast_\mu,\nonumber\\
{\cal B}_{\hat\Sigma}^{(0)}c^\ast&=&-\de^\mu\hat A^\ast_\mu,\nonumber\\
{\cal B}_{\hat\Sigma}^{(0)}\rho^\ast&=&\de^\mu\hat\eta^\ast_\mu.
\label{due.8}
\eea
with
\eq
{\cal B}_{\hat\Sigma}^{(0)}{\cal B}_{\hat\Sigma}^{(0)}=0.
\eqn{nilp-lin-approx}
The reason for looking at the operator ${\cal B}_{\hat\Sigma}^{(0)}$ relies on
a very general theorem on BRST cohomology stating that the cohomology of the
operator
${\cal B}_{\hat\Sigma}$ is isomorphic to a subspace of the cohomology of its
linearized approximation ${\cal B}_{\hat\Sigma}^{(0)}$.
\noindent
Making now the following linear change of variables
\bea
&&B_{\mu\nu}\rightarrow \tau_{\mu\nu}=B_{\mu\nu}-\de_{[\mu}\eta_{\nu]}
+{1\over{4g^2}}\varepsilon_{\mu\nu\rho\sigma}\de^{[\rho}A^{\sigma]},\nonumber\\
&&\psi_\mu\rightarrow\varphi_\mu=\psi_\mu+\de_\mu\rho,\nonumber\\
&&\hat A^\ast_\mu\rightarrow\hat\omega^\ast_\mu=\hat A^\ast_\mu-
{1\over{4g^2}}\varepsilon_{\mu\nu\rho\sigma}\de^\nu\hat
B^{\ast\rho\sigma},\nonumber\\
&&\hat\eta^\ast_\mu\rightarrow \hat\lambda^\ast_\mu=\hat\eta^\ast_\mu
+\de^\nu\hat B^\ast_{\mu\nu},\nonumber\\
&&\rho^\ast\rightarrow \xi^\ast=\rho^\ast+\de^\mu\hat\psi^\ast_\mu.
\label{due.9}
\eea
the other fields and sources remaining unchanged,
the action of ${\cal B}_{\hat\Sigma}^{(0)}$ can be written as
\bea
&&{\cal B}_{\hat\Sigma}^{(0)}A_\mu=-\de_\mu c,\nonumber\\
&&{\cal B}_{\hat\Sigma}^{(0)}c=0,\nonumber\\
&&{\cal B}_{\hat\Sigma}^{(0)}c^\ast= -\de^\mu\hat\omega^\ast_\mu, \nonumber\\
&&{\cal B}_{\hat\Sigma}^{(0)}\hat\omega^\ast_\mu=-{1\over{2g^2}}
\de^\nu\de_{[\mu}A_{\nu]}.
\label{ym-like}
\eea
and
\bea
&&{\cal B}_{\hat\Sigma}^{(0)}\hat B^\ast_{\mu\nu}=2g^2
\tau_{\mu\nu}, \qquad
 {\cal B}_{\hat\Sigma}^{(0)}\tau_{\mu\nu}=0,\nonumber\\
&&{\cal B}_{\hat\Sigma}^{(0)}\eta_\mu=\varphi_\mu, \qquad \qquad  
{\cal B}_{\hat\Sigma}^{(0)}\varphi_\mu=0,\nonumber\\
&&{\cal B}_{\hat\Sigma}^{(0)}\rho=-\phi, \qquad \qquad  
{\cal B}_{\hat\Sigma}^{(0)}\phi=0,\nonumber\\ 
&&{\cal B}_{\hat\Sigma}^{(0)}\hat\psi^\ast_\mu=
-\hat\lambda^\ast_\mu, \qquad \qquad
 {\cal B}_{\hat\Sigma}^{(0)}\hat\lambda^\ast_\mu=0,\nonumber\\
&&{\cal B}_{\hat\Sigma}^{(0)}\phi^\ast=-\xi^\ast, \qquad \qquad
{\cal B}_{\hat\Sigma}^{(0)}\xi^\ast=0.
\label{doublets}
\eea
From eqs.(\ref{doublets}) it is apparent that the variables
$(\tau,\hat B^\ast),(\varphi,\eta),(\phi,\rho),
(\hat\lambda^\ast,\hat\psi^\ast),(\xi^\ast,\phi^\ast)$ are grouped in BRST
doublets
\cite{ps} and therefore cannot contribute to the cohomology of
${\cal B}_{\hat\Sigma}^{(0)}$. We are left thus only with the fields and
sources appearing in the equations (\ref{ym-like}). However,
the latters are easily recognized to be nothing but the linearized
transformations characterizing the cohomolgy of the pure YM theory \cite{ps}.
This means that we can take as the representatives of the cohomolgy classes of
the operator
${\cal B}_{\hat\Sigma}$ those of the pure YM.

Therefore, the only nontrivial elements  (of canonical dimension 4 and ghost
number $1$ and $0$) of the cohomology of ${\cal B}_{\hat\Sigma}$ are given by
\begin{itemize}
\item
the usual nonabelian gauge anomaly
\be
{\cal A}=\varepsilon_{\mu\nu\rho\sigma}\int d^4x\: c_a\de^\mu
\left(d^{abc}\de^\nu A^\rho_b A^\sigma_c+{{{\cal D}^{abcd}}\over 12}
A^\nu_b A^\rho_c A^\sigma_d\right),
\label{due.11}
\ee
where $d_{abc}$ is the totally symmetric invariant tensor of rank 3 defined by
\be
d_{abc}={1\over 2}Tr(T_a\{T_b,T_c\}),
\label{due.12}
\ee
and
\be
{\cal D}_{abcd}=d^n_{ab}f_{ncd}+d^n_{ac}f_{ndb}+d^n_{ad}f_{nbc},
\label{due.13}
\ee
where $f_{abc}$ are the structure constants of the gauge group.
Since in our model all the fields are in the adjont representation,
the anomaly coefficient automathically vanishes to one loop order. The
Adler-Bardeen theorem guaranties thus that the gauge anomaly is definitively
absent to all
orders of perturbation theory.
\item
the most general invariant counterterm can be written as
\be
\Sigma^c=-{{Z_g}\over{4g^2}}\int d^4x\:\tr(F_{\mu\nu}F^{\mu\nu})
+{\cal B}_{\hat\Sigma}{\hat \Delta^{-1}},
\label{due.14}
\ee
where ${Z_g}$ is an arbitrary free parameter and $\Delta^{-1}$ is some local
integrated polynomial with ghots number -1 and dimension 4.
\end{itemize}
To a first look the counterterm (\ref{due.14}) does not seem to have the form
of the action $I_{BF-YM}$ of eq.\equ{class-action}. However, it is very easy
to check
that expression (\ref{due.14}) can be rewritten in the same form of the
original action
$I_{BF-YM}$, the difference being an irrelevant trivial cocycle, {\it i.e.}
\bea
&&{1\over{4g^2}}\int d^4x\:\tr(F_{\mu\nu}F^{\mu\nu})=2I_{BF-YM}
\nonumber\\
&&+{1\over{4g^2}}{\cal B}_{\hat\Sigma}\int d^4x\:\tr\bigg(\hat B^{\ast\mu\nu}
\left[{1\over 2}\varepsilon_{\mu\nu\rho\sigma}F^{\rho\sigma}
+2g^2(B_{\mu\nu}-D_{[\mu}\eta_{\nu]})^2\right]\bigg).
\label{due.15}
\eea
{}From eqs.(\ref{due.14}) and (\ref{due.15}) it then follows that there is only
one
physical renormalization, associated to the parameter ${Z_g}$, which gives a
non-vanishing $\beta_g$ for
the coupling constant $g$.
As expected, the numerical value of the 1-loop contribution to the $\beta_g$
function
is the same of the standard YM \cite{mz}.

Let us summarize our results.
We have found that the
BF-YM theory can be characterized in terms of the BRST cohomology of the pure
YM theory.
Moreover, the only non trivial counterterm (\ref{due.14})
can be equivalently rewritten in terms of the classical BF-YM action, 
yielding a renormalization of the gauge coupling $g$.

This shows the algebraic equivalence between the model described by the
classical action $I_{BF-YM}$ and the standard Yang-Mills theory.

These same conclusions are drawn in Ref.\cite{henn} which
appeared few hours before this paper was ready to be sent to
the publisher.  


\vskip 1.5cm
\leftline{\bf\large Acknowledgments}
\vskip .5cm
We thank P.Cotta-Ramusino e A.S.Cattaneo for discussions.


\end{document}